\documentclass{mem}
\usepackage{natbib}
\usepackage{txfonts}
\usepackage{balance}
\usepackage{graphicx}
\usepackage[a4paper]{hyperref}
\begin{document}
\def\teff{$T\rm_{eff }$}
\def\kms{$\mathrm {km s}^{-1}$}

\title{Concluding Remarks}

\author{Tim de Zeeuw\inst{1}}

\offprints{P.T.\ de Zeeuw}
 
\institute{Sterrewacht, Leiden University, Niels Bohrweg 2, 2333 CA
           Leiden, The Netherlands \\ 
           \email{dezeeuw@strw.leidenuniv.nl} }

\authorrunning{de Zeeuw}

\titlerunning{Concluding Remarks}

\abstract{Summary of Joint Discussion 13 of the 26th IAU General Assembly, 
          Prague.}


\maketitle{}

\noindent
It is a pleasure to thank the organizers for putting together such an
interesting and enjoyable Joint Discussion on exploiting ongoing and
future large surveys for galactic astronomy. The program covered a
little over one day, and concluded with a wide-ranging discussion
session, involving many of the speakers and other participants. It was
good, but also slightly nerve-wrecking, to hear many of the points I
had planned to mention coming up naturally in the past hour.  For this
reason I will restrict myself here to a brief overview of the key
science goals and some highlights, and then give my personal view on
the challenges and opportunities for the future.\looseness=-2

\section{Galaxy formation \& the Milky Way}

Much world-wide observational effort concentrates on understanding
galaxy formation. It uses the principle that `far away $=$ long ago',
which is sometimes referred to as the fundamental equation of
astronomy. Numerical simulations of structure formation by now have
the resolution to model individual galaxies. The motions and
properties of old stars contain much of the `fossil record' of the
formation history. This fossil record can be read in detail in the
Milky Way and its nearest neighbors, because of our ability to resolve
the individual stars, and measure their properties. The Milky Way is
an average galaxy, in a representative Local Group, so understanding
how it formed is central to the entire field of Galaxy formation. This
`near-field' cosmology \citep{fbh02} provides a key test of the
paradigm of formation through hierarchical merging of many small
building blocks, as summarized by Wyse, Bland--Hawthorn, and
Helmi. Specifically, we need to determine i) When the stars in the
Milky Way formed, ii) When and how the Milky Way was assembled, and
iii) How dark matter in the Milky Way is distributed. It is no
coincidence that these are exactly the top three science goals for the
stereoscopic census of the Milky Way to be provided by Gaia
\citep{pea01}, and indeed for many ongoing and planned
surveys.\looseness=-2

\section{Surveys from ground and space}

Many exciting photometric surveys are being carried out, including
2MASS and SDSS/SEGUE (Newberg, Ivezi\'c). They reveal many coherent
structures in the Galactic halo which include tidal tails of globular
clusters and debris from accreted satellites. Akari is carrying out a
mid-infrared all-sky survey (Ishihara), and both VST/VISTA and
Pan-Starrs 1 (Kaiser) are about to come on line. Ambitious plans for
massive follow-up with Pan-Starrs 4 and LSST are in place, but these
are not yet funded.

Many thousands to multiple millions of radial velocities are being
provided by the Geneva/Copenhagen survey (Holmberg), SDSS/SEGUE
(Ivezi\'c), the RAVE project (Steinmetz) and, in the future, by
Gaia. The tremendous scientific impact of the HIPPARCOS proper motions
and parallaxes for the brightest 120000 stars has stimulated much
astrometric work from the ground, using digitized old photographic
plates and new surveys with large-area CCD detectors (Turon). A major
jump in accuracy will come when ESA launches Gaia in 2011, which will
provide proper motions and parallaxes for a billion stars to magnitude
20. Institutions in Japan and the US have plans for other space
astrometry missions as well.

Many of the photometric and kinematic surveys also measure (or at
least estimate) stellar parameters including $T_{\rm eff}$, $\log g$,
the extinction $A_V$, [Fe/H], and [$\alpha$/Fe]. High-resolution
spectroscopy for subsets of survey stars has already provided more
accurate values, as well as specific elemental abundances (Feltzing).

I was impressed by the mounting evidence for different kinematic
groups with very homogeneous elemental abundances within the group,
but distinct differences between groups. Such evidence is found not
only in the stellar halo, but now also in the thick disk (Feltzing,
Helmi), and is very suggestive of successive merging of separate
building blocks. Helmi demonstrated convincingly that these building
blocks are not the same as the precursors of the dwarf spheroidals
that we observe around the Milky Way today.

All of this shows that systematic kinematic selection from unbiased
large-scale samples, in particular the dataset to be provided by Gaia,
and follow-up high-resolution spectroscopy, holds much promise for
unraveling at least part of the formation history of the Local
Group. This can be done in unprecedented detail in the Milky Way, and
is also becoming possible for the nearest galaxies with instruments
such as {\tt FLAMES} on the VLT (Helmi).

\section{Challenges}

It became clear during the presentations that the ongoing and future
large survey projects pose some significant challenges. It is critical
to calibrate the various groundbased photometric surveys (Glass), and
to be very careful about the derivation of stellar parameters from
photometry or spectroscopy (Gray). Methods that have been developed
for some class of stars should not be applied blindly to all types.

The importance of correcting for extinction was stressed by some
speakers, and ignored by others. It may not be an issue for halo stars
at substantial galactic latitudes, but correcting for extinction will
be crucial for planned studies of the disk and the Bulge.  Galaxy
models can help (Robin), as do surveys of the atomic and molecular
gas, and studies in the infrared (2MASS, VISTA), but there is little
substitute for measuring the distances independently, via the
parallax.  Gaia will provide the major step forward here, but it is
critical that ESA now holds firm on the astrometric specifications.

Variable stars, including eclipsing binaries, can provide much
additional physical information on stellar properties and distances,
but this requires special attention, including a well-planned cadence
of multi-epoch observations (Cook). There is little freedom to do this
with Gaia, as it will spin at a fixed rate, but ample opportunity for
the groundbased surveys.

Much general preparation is needed to extract the key scientific
results from the large databases generated by surveys. For example, it
will be very interesting to go beyond star counts in cells, and to
connect them with the observed kinematics by use of Newton's laws of
motion and gravitation. When applied to the full Gaia data, this will
provide massive discriminating power between models and formation
scenarios. A number of approaches are known for carrying out the
required dynamical modeling, but the existing machinery needs to be
developed significantly before it can deal with the large data sets of
the future. Work on this should start now. Much experience is
available in various institutes in Europe, and support by the national
funding agencies, or by an EU RTN, would be very helpful.

\section{Opportunities} 

There is much complementarity between the various ground-based surveys
and the data to be expected from the Gaia all-sky survey
(Bailer-Jones). The RAVE effort is an excellent example. It gives
exciting scientific results for the brighter stars, and also provides
pilot experience with precisely the kind of spectroscopic data to be
expected from Gaia for perhaps a 100 million objects fainter than
magnitude 12.

The Gaia photometry will result in accurate and homogeneous color
information for a billion objects to magnitude 20 over the entire
sky. SDSS already extends this a few magnitudes fainter, over a large
sky area, and the plan is that VST/VISTA, Pan-Starrs 4 and LSST would
go fainter still (Kaiser), and even provide some modest accuracy
proper motions beyond the Gaia limit. The Gaia space photometry will
be very helpful to calibrate all the groundbased measurements, so that
they can be used with confidence at these faint
magnitudes.\looseness=-2

The Japanese space astrometry plans are exciting as they focus on an
infrared study of the Bulge, which cannot be reached by Gaia other
than in low-extinction windows. NASA's SIM will focus on exoplanets,
as it should, but may provide extremely accurate parallaxes and proper
motions for a modest set of pre-selected stars. I am puzzled by the
OBSS mission proposed by the US Naval Observatory. The science aims
appear copied from Gaia, but would be achieved through a different
technical solution. It might be good to have a backup in case of a
mishap with Gaia, but the scope of the space science ambitions of NASA
and ESA relative to their budgets does not seem to favor
trans-atlantic duplication. Surely a better way forward could be
explored.

Finally, the proposed {\tt WFMOS} for Subaru with funding from the
International Gemini Observatory would provide a marvellous
opportunity for the needed high-resolution spectroscopic follow-up. It
was not clear from the talks whether even more ambitious instruments
or dedicated telescopes would be required. If so, then development
better start soon.

\section{Conclusions}

The next decade will see tremendous progress in our understanding of
the formation of the Milky Way and the Local Group. Homogeneous large
scale photometric, radial velocity and astrometric surveys, dedicated
high-resolution spectroscopic follow-up, and theoretical preparation
are critical. The talks and the discussion showed that the
complementarity of surveys from ground and from space can probably be
exploited further with the Gaia mission as the centerpiece, and that
plans for this are underway. There is much to look forward to.

\bibliographystyle{aa}

\begin{thebibliography}{}

\bibitem[{Freeman \& Bland--Hawthorn (2002)}]{fbh02}
Freeman, K.\ C., Bland--Hawthorn, J. 2002, ARAA, 40, 487

\bibitem[{Perryman et~al.\ (2001)}]{pea01}
Perryman, M.~A.~C., de Boer, K.~S., Gilmore, G., 
H{\o}g, E., Lattanzi, M.~G., Lindegren, L., Luri, X., 
Mignard, F., Pace, O., de Zeeuw, P.~T. 2001, \aap, 369, 339

\end{thebibliography}

\end{document}